\documentclass[conference, 10pt]{IEEEtran}
\IEEEoverridecommandlockouts

\usepackage{amsmath,amsfonts,bm}

\def\eqref#1{equation~\ref{#1}}

\def\1{\bm{1}}

\def\vtheta{{\bm{\theta}}}
\def\va{{\bm{a}}}
\def\vb{{\bm{b}}}

\DeclareMathAlphabet{\mathsfit}{\encodingdefault}{\sfdefault}{m}{sl}
\SetMathAlphabet{\mathsfit}{bold}{\encodingdefault}{\sfdefault}{bx}{n}

\usepackage{cite}
\usepackage{amsmath,amssymb,amsfonts}
\usepackage{algpseudocode}
\usepackage{graphicx}
\usepackage{textcomp}
\usepackage{xcolor}
\usepackage{braket}
\usepackage{url}
\usepackage{hyperref}
\usepackage{booktabs} 
\usepackage{multirow} 

\hypersetup{
  colorlinks   = true, 
  urlcolor     = blue, 
  linkcolor    = blue, 
  citecolor   = blue 
}

\def\BibTeX{{\rm B\kern-.05em{\sc i\kern-.025em b}\kern-.08em
    T\kern-.1667em\lower.7ex\hbox{E}\kern-.125emX}}
\begin{document}

\title{Quantum-Enhanced Parameter-Efficient Learning for Typhoon Trajectory Forecasting	 \thanks{The views expressed in this article are those of the authors and do not represent the views of Wells Fargo. This article is for informational purposes only. Nothing contained in this article should be construed as investment advice. Wells Fargo makes no express or implied warranties and expressly disclaims all legal, tax, and accounting implications related to this article.}
}


\author{
\IEEEauthorblockN{
     Chen-Yu Liu \IEEEauthorrefmark{1}\IEEEauthorrefmark{6}, 
     Kuan-Cheng Chen\IEEEauthorrefmark{2}\IEEEauthorrefmark{3}\IEEEauthorrefmark{7},
    Yi-Chien Chen \IEEEauthorrefmark{4} \\
    Samuel Yen-Chi Chen\IEEEauthorrefmark{5}, 
    Wei-Hao Huang \IEEEauthorrefmark{12},
    Wei-Jia Huang\IEEEauthorrefmark{9}, 
    Yen-Jui Chang \IEEEauthorrefmark{10}\IEEEauthorrefmark{11}
}

\IEEEauthorblockA{\IEEEauthorrefmark{1}Graduate Institute of Applied Physics, National Taiwan University, Taipei, Taiwan}
\IEEEauthorblockA{\IEEEauthorrefmark{9}Hon Hai (Foxconn) Research Institute, Taipei, Taiwan}
\IEEEauthorblockA{\IEEEauthorrefmark{2}Department of Electrical and Electronic Engineering, Imperial College London, London, UK}
\IEEEauthorblockA{\IEEEauthorrefmark{3}Centre for Quantum Engineering, Science and Technology (QuEST), Imperial College London, London, UK}
\IEEEauthorblockA{\IEEEauthorrefmark{4} Institute for Space-Earth Environmental Research, Nagoya University, Nagoya, Japan}
\IEEEauthorblockA{\IEEEauthorrefmark{5}Wells Fargo, New York, NY, USA}
\IEEEauthorblockA{\IEEEauthorrefmark{12}Jij Inc., Tokyo, Japan}
\IEEEauthorblockA{\IEEEauthorrefmark{10}Quantum Information Center, Chung Yuan Christian University, Taoyuan City, Taiwan}
\IEEEauthorblockA{\IEEEauthorrefmark{11}Master Program in Intelligent Computing and Big Data, Chung Yuan Christian University, Taoyuan City, Taiwan}
\IEEEauthorblockA{Email:\IEEEauthorrefmark{6} d10245003@g.ntu.edu.tw, \IEEEauthorrefmark{7}kuan-cheng.chen17@imperial.ac.uk}

}

\maketitle
\begin{abstract}
Typhoon trajectory forecasting is essential for disaster preparedness but remains computationally demanding due to the complexity of atmospheric dynamics and the resource requirements of deep learning models. Quantum-Train (QT), a hybrid quantum-classical framework that leverages quantum neural networks (QNNs) to generate trainable parameters exclusively during training, eliminating the need for quantum hardware at inference time. Building on QT’s success across multiple domains, including image classification, reinforcement learning, flood prediction, and large language model (LLM) fine-tuning, we introduce Quantum Parameter Adaptation (QPA) for efficient typhoon forecasting model learning. Integrated with an Attention-based Multi-ConvGRU model, QPA enables parameter-efficient training while maintaining predictive accuracy. This work represents the first application of quantum machine learning (QML) to large-scale typhoon trajectory prediction, offering a scalable and energy-efficient approach to climate modeling. Our results demonstrate that QPA significantly reduces the number of trainable parameters while preserving performance, making high-performance forecasting more accessible and sustainable through hybrid quantum-classical learning.

\end{abstract}


\section{Introduction}
Climate change has intensified the frequency and severity of typhoons, leading to devastating consequences for communities worldwide. The increased frequency of extreme weather events underscores the urgent need for more accurate trajectory forecasts to mitigate human and economic losses \cite{emanuel2005increasing, webster2005changes, ipcc2021report}. Accurate forecasting allows for timely disaster response, evacuation planning, and resource allocation, significantly reducing casualties and infrastructure damage. Taiwan, characterized by its steep and mountainous terrain, is particularly vulnerable to extreme weather events. On average, the island experiences approximately 3.5 typhoons and dozens of torrential rainstorms each year, resulting in annual economic losses of \textbf{374.3 million} Euros due to infrastructure damage, agricultural losses, and disruptions to economic activity \cite{nhess-2022-38}. The high frequency and intensity of these events highlight the critical need for advanced forecasting models to improve disaster preparedness and resilience.

Traditional computational approaches to weather prediction have made substantial progress \cite{bauer2015quiet}, but they often struggle to model the inherent complexity of typhoon dynamics. These challenges arise due to the chaotic nature of atmospheric systems, the high-dimensional parameter spaces, and the need for extensive computational resources to process vast amounts of meteorological data. Current numerical weather prediction (NWP) models rely on large-scale supercomputing infrastructure, which is costly and energy-intensive. As both data volume and model sophistication continue to increase, these limitations become more pronounced, necessitating more efficient and scalable forecasting methodologies.

A key challenge in modern weather forecasting is the high computational cost associated with training complex models. Large-scale meteorological models require extensive parameter tuning and optimization, leading to:
\begin{itemize}
\item \textbf{High energy consumption:} Training deep learning models for weather forecasting demands significant computing power, contributing to high operational costs and carbon emissions.
\item \textbf{Limited accessibility:} Developing countries and resource-constrained research institutions may lack access to the supercomputing infrastructure needed for training state-of-the-art forecasting models.
\item \textbf{Scalability concerns:} As weather models grow in complexity, their training becomes increasingly resource-intensive, limiting their deployment for real-time forecasting applications.
\end{itemize}

Thus, parameter-efficient training methods are essential for enhancing the sustainability, accessibility, and scalability of advanced weather forecasting models. Reducing the number of trainable parameters while maintaining model accuracy provides several advantages: (1) it accelerates model convergence, enabling faster updates and adaptation to evolving weather patterns; and (2) it improves generalization, as efficient models are less susceptible to overfitting, thereby enhancing resilience to shifting climate conditions.

Quantum computing emerges as a promising paradigm that can address high-dimensional optimization and state-space exploration with unprecedented efficiency \cite{montanaro2016quantum}. By integrating quantum-enhanced methods with classical computing—an approach termed quantum-centric supercomputing \cite{bravyi2022future, gambetta2022quantum}—these complementary strengths alleviate computational bottlenecks. Decomposing complex problems into components suited for classical or quantum computing allows for more efficient problem-solving schemes. Classical systems are well-suited for tasks like data processing, while quantum computing shows potential in optimization and exploring large state spaces \cite{chen2024noise, liu2023hybrid, booth2017partitioning, phillipson2021portfolio, liu2024quantumlocal, raymond2023hybrid, liu2022implementation,hsu2024quantumklstm,hsu2024quantum,ma2025robust, chen2025learning,chen2024consensus,chen2024compressedmediq,chen2025resource,chen2024validating}.

Conventional quantum machine learning (QML) approaches employ parameterized quantum circuits (PQCs) as quantum neural networks (QNNs) \cite{chen2020variational}, where data is input through specific data encoding methods \cite{perez2020data, schuld2021effect}. The updates to QNN parameters during the training process are computed on the classical side, creating a hybrid quantum-classical computing framework \cite{mari2020transfer}. While QML offers improvements in specific applications \cite{cerezo2022challenges, huang2022quantum, biamonte2017quantum, caro2022generalization, chen2024quantum, huang2021power, chen2025toward}, significant challenges remain—particularly in data encoding for large datasets.

One such proposal to address both the data encoding challenge and the requirement for quantum hardware during the inference stage is Quantum-Train (QT) \cite{liu2024quantum, liu2024introduction}. Instead of using a QNN to interact directly with the data, this approach leverages QNNs to generate the weights of a target classical neural network model during the training process. The QT method:
\textbf{\textcolor{blue}{(1)} Keeps the data input process entirely within the classical model. \textcolor{blue}{(2)} Eliminates the need for quantum computing hardware during inference, as the trained model is purely classical. \textcolor{blue}{(3)} Reduces the number of training parameters on a polylogarithmic scale by mapping quantum state basis to the target neural network parameters ($M \rightarrow \mathrm{polylog}(M)$)}. Current studies on QT have demonstrated promising results across various domains, including image classification \cite{liu2024quantum, lin2024quantum2}, reinforcement learning \cite{liu2024qtrl, chen2024quantum2,chen2024DQTRL}, long short-term memory (LSTM) models and regression tasks \cite{lin2024quantum, liu2025federated, liu2024programming}.  
Notably, the LSTM-based flood prediction work won second place in the Deloitte's Quantum Climate Challenge 2024\footnote{Deloitte's Quantum Climate Challenge 2024: \url{https://www.deloitte.com/de/de/issues/sustainability-climate/deloitte-quantum-climate-challenge.html}}. QT effectively reduces the number of trainable parameters while maintaining task performance. Importantly, the trained model remains fully classical, enabling deployment on classical computers without requiring quantum hardware for inference.

As a further extension from QT, Quantum Parameter Adaptation (QPA) is proposed \cite{liu2025a}. QPA utilize the concept of QT—leverage the Hilbert space to do the parameter compression—but further extend the concept from directly compressing the machine learning (ML) model to compressing the parameter-efficient fine-tuning parameters \cite{hu2021lora, liu2024dora, houlsby2019parameter, lin2020exploring, li2021prefix, yang2021voice2series}. That is, consider a low-rank adaptation (LoRA) for a large language model (LLM), two low-rank matrices $A \in \mathbb{R}^{l \times r}$ and $B \in \mathbb{R}^{r \times k}$ are generated to represent the update $AB =\Delta W  \in \mathbb{R}^{l \times k}$ of a given weight matrix $W \in \mathbb{R}^{l \times k}$, with the rank $r \ll \min(l,k)$. QPA uses QT to generate the parameters of $A$ and $B$, thus it is possible to use extremely small amount of parameters (in the scale of $\mathrm{polylog}(lr+rk)$) to fine-tune LLMs. 


Drawing inspiration from these successes—particularly in flood prediction and LLM fine-tuning—we propose extending QT to tackle climate challenges that require large ML models, specifically typhoon trajectory forecasting. This task is known for its high computational demands, yet we believe that the latest advancements in QT offer a more efficient and scalable solution.

\section{Quantum Approach for Classical Parameter-Efficient Learning Tasks}

\subsection{Quantum-Train}

To introduce the QT method, we begin by considering a neural network (NN) defined for a specific machine learning task. At this point, we are not concerned with the exact architecture of the NN; instead, we treat its trainable parameters as a vector of size $m$, $\va = (a_1, a_2, \hdots, a_m)$, with each entry representing a weight or bias to be optimized.

The QT framework diverges from traditional QML approaches. Instead of encoding input data into quantum states or using quantum layers within classical models, QT uses a PQC to indirectly guide the training of a classical NN. The PQC is composed of $N = \lceil \log_2 m \rceil$ qubits and $L$ layers, with the circuit constructed from standard building blocks such as rotation gates and controlled-NOT (CNOT) gates. A typical ansatz used in QT is defined as follows:
\begin{equation}
\label{eq:ansatz}
|\psi(\vtheta)\rangle = \left( \prod_{i=1}^{N-1} \text{CNOT}^{i, i+1} \prod_{j=1}^{N} R_Y^j (\theta_{j}^{(L)}) \right)^L |0\rangle^{\otimes N},
\end{equation}
where $R_Y^j(\theta)$ is a single-qubit rotation gate around the $Y$-axis applied to qubit $j$ with tunable angle $\theta_{j}^{(L)}$, and $\text{CNOT}^{i, i+1}$ denotes a CNOT gate between qubits $i$ and $i+1$. The number of layers $L$ functions as a hyperparameter and typically scales with $N$, for example, $L = O(N)$ or $L = O(N^2)$, though more generally, one can consider $L = O(\mathrm{poly}(N))$.

Executing this PQC yields a quantum state from which $2^N$ measurement probabilities are obtained. These values, lying in the interval $[0, 1]$, correspond to the probabilities of observing each computational basis state. Since $2^N \geq m$, there are enough distinct outcomes to cover all the parameters of the target NN.

To convert these measurement results into usable NN parameters, we introduce a lightweight classical mapping model $G_\vb$, such as a multilayer perceptron (MLP). The input to this model is a combination of the binary representation of a basis state (of length $N$) and its associated measurement probability. The output is the corresponding parameter value $a_i$ for the NN. In other words,
\[
G_\vb\left(|\phi_i\rangle, |\langle \phi_i | \psi(\vtheta) \rangle|^2\right) = a_i, \quad \forall i \in \{1, 2, \ldots, m\}.
\]
Here, only the first $m$ basis states are considered to construct the full parameter vector. Since the input to $G_\vb$ is of size $N+1$, its number of trainable parameters can scales as $O(\mathrm{poly}(N))$, or equivalently $O(\mathrm{polylog}(m))$, depends on the design of the MLP architecture.

\begin{figure*}[h]
\centering
\includegraphics[width=0.9\linewidth]{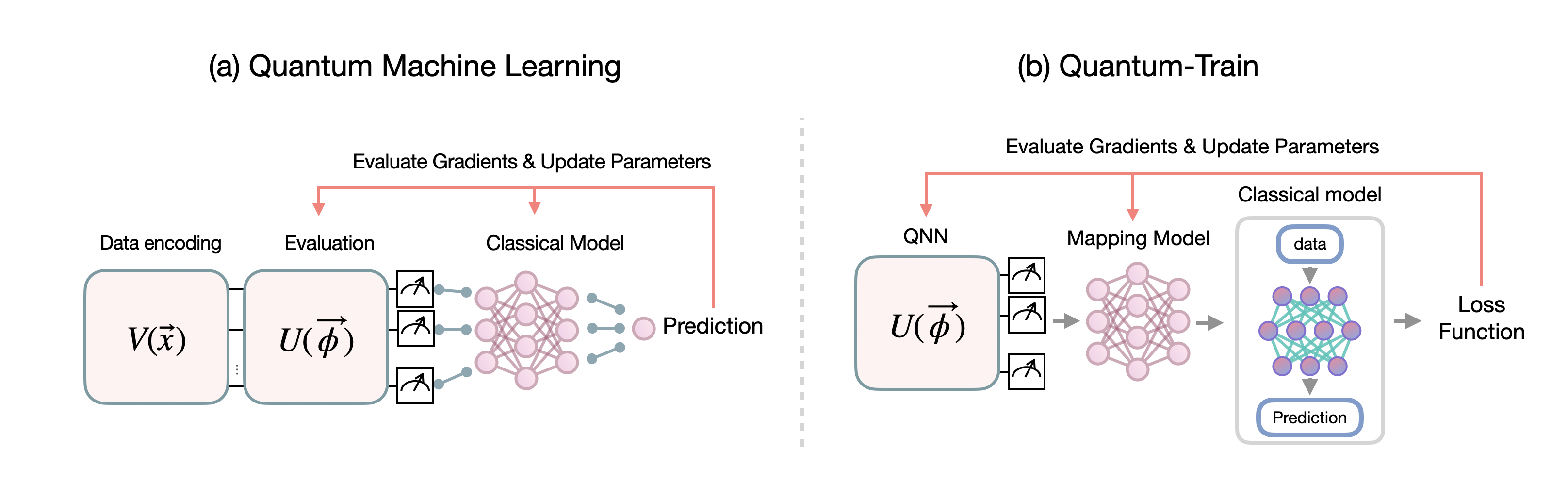}
\caption{
Overview of (a) Quantum Machine Learning \cite{mari2020transfer, mitarai2018quantum}, (b) Quantum-Train \cite{liu2024quantum}
}
 \label{fig:qml_and_qt}
\end{figure*}

The output vector $\va$ generated through this process is then used to define a classical NN for the given learning task. During training, both the PQC parameters $\vtheta$ and the mapping model parameters $\vb$ are updated based on a classical loss function $\mathcal{L}$, such as the mean square error (MSE) loss in regression tasks. This hybrid optimization procedure enables the training of a large NN using a significantly smaller set of trainable parameters, distributed across both quantum and classical components. A high-level comparison between this QT-based workflow and conventional QML approaches is illustrated in Fig.~\ref{fig:qml_and_qt}.

\textbf{\textit{Gradient Estimation for Quantum-Compressed Parameters.}} 
In the QT framework, the target NN parameters $\va$ are generated through a combination of a PQC and a classical mapping model. The quantum-influenced variables, collectively denoted as $(\vtheta, \vb)$, govern the generation of $\va$ through state preparation and measurement processes. The gradient of the loss function with respect to these quantum parameters can be expressed as:

\begin{equation}
\label{eq:grad}
\nabla_{\vtheta,\vb} \mathcal{L} = \left( \frac{\partial \va}{\partial (\vtheta,\vb)} \right)^T \cdot \nabla_{\va} \mathcal{L}.
\end{equation}

Here, the term $\frac{\partial \va}{\partial (\vtheta,\vb)}$ represents the Jacobian matrix, capturing the sensitivity of the generated classical parameters $\va$ to variations in the underlying quantum-train parameters $(\vtheta, \vb)$. This formulation provides a clear and differentiable pathway for optimization when exact quantum state simulation is used. In experimental settings involving real quantum hardware, gradient evaluation must rely on techniques such as the parameter-shift rule and its extensions \cite{mitarai2018quantum, schuld2019evaluating} to estimate derivatives efficiently under finite sampling conditions.

\textbf{\textit{Parameter Update for Quantum-Compressed Models.}} 
The learning rate $\eta$ plays a crucial role in guiding the optimization process, particularly given the hybrid quantum-classical nature of the system. The quantum parameters are updated using a first-order optimization rule, given by:

\begin{equation}
\label{eq:update}
\vtheta_{t+1}, \vb_{t+1} = \vtheta_t, \vb_t + \eta \nabla_{\vtheta, \vb} \mathcal{L}.
\end{equation}

This update mechanism ensures that the quantum circuit parameters and the mapping model are jointly refined to improve the overall performance of the target neural network.

Using the gradient computation and update rules described above, the QT framework has been applied successfully across a variety of domains as mentioned in the introduction. Across these applications, QT has consistently demonstrated a substantial reduction in trainable parameter counts, while achieving competitive performance. This highlights QT’s potential as a broadly applicable strategy for parameter-efficient learning that leverages the unique advantages of quantum resources during training.

\textbf{\textit{Batched Parameter Generation Strategy.}} 
In scenarios where the target NN contains $m$ trainable parameters, the number of qubits required by the QT framework is given by $N = \lceil \log_2 m \rceil$. For instance, scaling up to $m = 10^9$ parameters (one billion) would imply a need for $N = 30$ qubits. Although constructing a quantum circuit with 30 qubits is feasible on current quantum hardware and classical simulators, such configurations impose heavy demands on computational resources—approximately 16 GB of GPU memory and several seconds per circuit evaluation \cite{google_qsim} for intermediate-level hardware. These costs can be prohibitive for ML workflows that require rapid, repeated iterations.

To mitigate this bottleneck, we adopt a batched parameter generation approach introduced in \cite{liu2024federated}. This method partitions the $m$ parameters of the NN into $n_{ch}$ chunks, each containing $n_{mlp}$ parameters, such that $n_{ch} = \lceil m / n_{mlp} \rceil$. By batching multiple NN parameters into a single output, we reduce the number of distinct quantum measurements required, and consequently, the number of qubits needed.

In this scheme, the mapping model $\Tilde{G}_{\vb}$ takes a quantum basis state $|\phi_i\rangle$ and its associated measurement probability $|\langle \phi_i | \psi (\vtheta) \rangle|^2$ as inputs, and outputs a batch of $n_{mlp}$ classical parameters:

\begin{eqnarray}
&& \va = (\Tilde{a}_1, \Tilde{a}_2, \hdots, \Tilde{a}_{n_{ch}}), \\
\label{eq:mapping_batch_1}
&& \Tilde{G}_{\vb} (|\phi_i \rangle, |\langle \phi_i | \psi (\vtheta) \rangle|^2) = \Tilde{a}_i, \quad \forall i \in \{ 1, 2, \ldots, n_{ch} \}, \\
\label{eq:mapping_batch_2}
&& \Tilde{a}_i = ( a_{i,1}, a_{i,2}, \dots, a_{i,j} ), \quad \forall j \in \{ 1, 2, \ldots, n_{mlp} \}.
\end{eqnarray}

This expansion is achieved using a decoder-style MLP architecture within $\Tilde{G}_{\vb}$, which increases the output dimensionality from one to $n_{mlp}$ per quantum sample. The architectural configuration used in this work is detailed in Table~\ref{tab:mm}. As a result, the required number of qubits is reduced from $N = \lceil \log_2 m \rceil$ to:

\begin{equation}
\label{eq:QTBG_qubits_usage}
N = \lceil \log_2 n_{ch} \rceil = \lceil \log_2 \left(  \lceil \frac{m}{n_{mlp}} \rceil \right) \rceil.
\end{equation}

This formulation effectively decreases the qubit count by approximately $\lceil \log_2 n_{mlp} \rceil$ relative to the original QT method (since $\log(\frac{a}{b}) = \log (a) - \log(b)$). Notably, the original QT setup can be viewed as a special case where $n_{mlp} = 1$. While this batching strategy leads to an increase in the parameter size of the mapping model due to its larger output layer, it delivers substantial memory and runtime benefits. Specifically, the memory footprint for quantum state storage is reduced by a factor of $1 / n_{mlp}$.

For example, if $m = 10^9$ and $n_{mlp} = 1024$, then the number of required qubits drops to:

\begin{equation}
N = \lceil \log_2 \left( \left\lceil \frac{10^9}{1024} \right\rceil \right) \rceil = 20.
\end{equation}

This reduction from 30 to 20 qubits represents a $33\%$ decrease in qubit requirements and yields a $1024$-fold reduction in memory usage for the quantum state. Such optimization significantly improves the practicality of classical simulation and hardware execution within the QT framework.

\begin{table*} 
    \centering
    \caption{Configuration of the mapping model $\Tilde{G}_{\vb}$, with $N$ representing the number of qubits for each task.}
    \vspace{10pt} 
    \small
    \begin{tabular}{llcc}
        \toprule
          \textbf{Hyperparameter} & \textbf{Meaning} & \textbf{Value} \\
        \midrule
         Input size & Input of the mapping model $(| \phi_i \rangle, |\langle \phi_i | \psi (\vtheta) \rangle|^2)$ & $N+1$  \\
         Hidden dimension & Main structure of the MLP mapping model & [32, 32, $n_{mlp}$]  \\
        \bottomrule
    \end{tabular}
    \label{tab:mm}
\end{table*}

\subsection{Overview of Parameter-Efficient Fine-Tuning Methods}

Before introducing the QPA approach, it is helpful to first review the concept of parameter-efficient fine-tuning (PEFT), which serves as a foundation and motivation for our method.

PEFT techniques aim to address the significant computational demands associated with fine-tuning LLMs. These methods focus on reducing the number of trainable parameters, often by several orders of magnitude, while retaining—if not improving—the model's performance on downstream tasks. One widely adopted strategy involves low-rank adaptation, as seen in methods such as LoRA \cite{hu2021lora} and DoRA \cite{liu2024dora}. These approaches are based on the insight that weight updates during fine-tuning typically lie in a low-rank subspace. Instead of adjusting the entire weight matrices, they introduce small, trainable low-rank matrices that, when composed with the original weights, capture the necessary task-specific adjustments. This dramatically reduces the number of parameters that must be optimized.

A different family of PEFT approaches utilizes \textit{adapters}—compact feed-forward modules inserted between the layers of a pretrained model \cite{houlsby2019parameter, lin2020exploring}. During fine-tuning, only the parameters within the adapter modules are updated, while the backbone model remains fixed. This strategy not only limits the memory and compute requirements but also facilitates modularity and rapid task adaptation.

Another notable technique is \textit{Prefix Tuning (PT)} \cite{li2021prefix, yang2021voice2series}, which departs from architectural modifications by prepending learnable vectors—called prefixes—to the input representations or intermediate hidden states within each layer of the model. Only these prefix vectors are updated during training, leaving the pretrained model weights entirely untouched. 

By concentrating the optimization on a small subset of parameters, PEFT methods effectively decouple task-specific adaptation from full model retraining. This makes them particularly well-suited for scalable fine-tuning of large models across multiple tasks or domains, and serves as a conceptual precedent for the resource-efficient principles adopted in our quantum-classical hybrid adaptation strategy.

\subsection{Quantum Parameter Adaptation}

Beyond the batched parameter generation strategy, a fundamental distinction between the original QT framework and the Quantum Parameter Adaptation (QPA) lies in the definition of the training objective. In QPA, the goal shifts from learning the complete set of NN parameters to learning only the parameters associated with a PEFT method. This adjustment significantly enhances scalability, making QPA well-suited for large-scale models and tasks. When combined with the batched parameter generation mechanism, QPA offers further improvements in computational efficiency and resource management.

Within this framework, the parameter vector $\va$ is redefined to represent the tunable parameters of a chosen PEFT technique. Consider, for instance, the Low-Rank Adaptation (LoRA) method. Given a pretrained weight matrix $W_0 \in \mathbb{R}^{d \times k}$, LoRA formulates the parameter update as a low-rank decomposition:

\[
W_0 + \Delta W = W_0 + BA,
\]

where $B \in \mathbb{R}^{d \times r}$ and $A \in \mathbb{R}^{r \times k}$, with the rank $r \ll \min(d, k)$. In QPA, these low-rank matrices $A$ and $B$ are generated using the same quantum-classical mapping procedure described previously in Eq.~\ref{eq:mapping_batch_1} and Eq.~\ref{eq:mapping_batch_2}, with $\va$ encoding the flattened entries of both $A$ and $B$.

The number of required qubits in this LoRA-based QPA setup is given by:

\begin{equation}
N = \lceil \log_2 \left( \left\lceil \frac{r(d + k)}{n_{mlp}} \right\rceil \right) \rceil.
\end{equation}

To illustrate the efficiency gains, suppose $W_0 \in \mathbb{R}^{2048 \times 1024}$, yielding over two million parameters in the full matrix. If we set $r = 4$, the low-rank factors become $B \in \mathbb{R}^{2048 \times 4}$ and $A \in \mathbb{R}^{4 \times 1024}$, totaling $r(d + k) = 12288$ parameters. With a chunk size of $n_{mlp} = 64$, the required number of qubits becomes:

\begin{equation}
N = \lceil \log_2 \left( \left\lceil \frac{4(2048 + 1024)}{64} \right\rceil \right) \rceil = 8.
\end{equation}

By contrast, applying the standard QT method directly to the full parameter matrix would require 21 qubits. This demonstrates a significant reduction in quantum resource requirements, highlighting the efficiency of QPA when applied to PEFT methods like LoRA.

\begin{figure*} 
\centering
\includegraphics[width=0.9\linewidth]{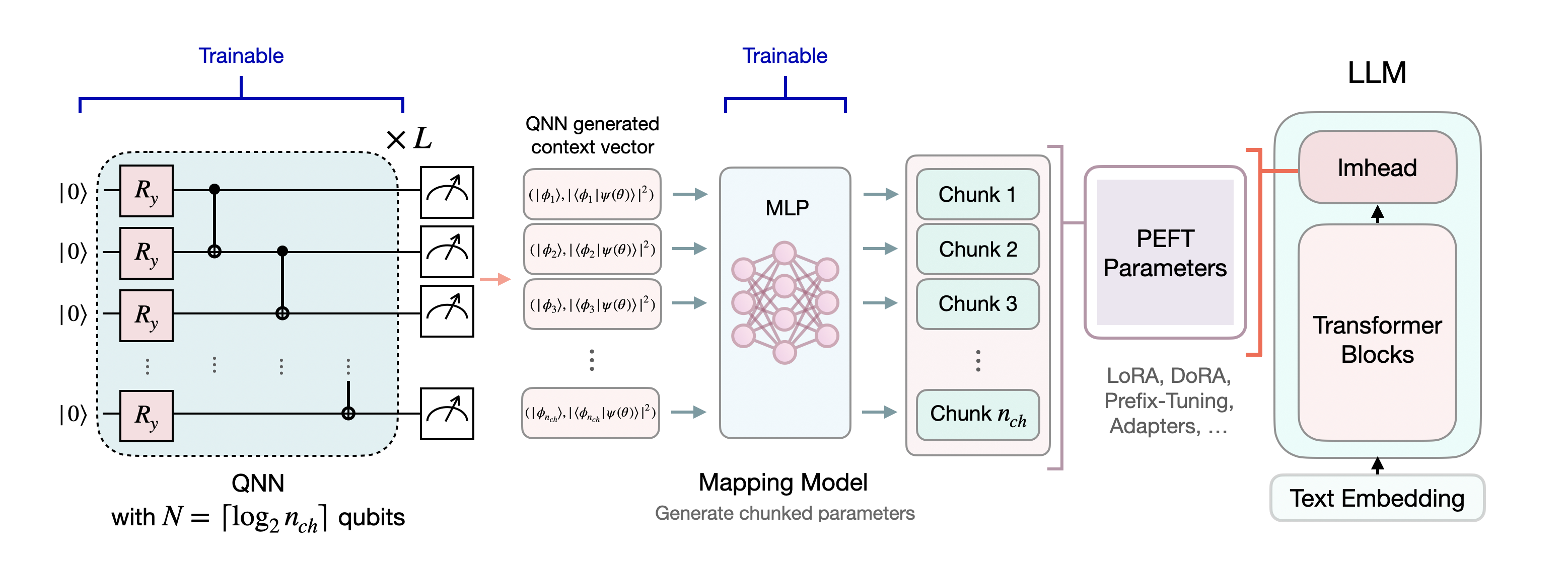}
\caption{
Overview of Quantum Parameter Adaptation \cite{liu2025a}.
}
 \label{fig:qpa}
\end{figure*}

The training process in QPA follows the same optimization dynamics introduced in Eq.~\ref{eq:grad} and Eq.~\ref{eq:update}, leveraging gradient-based methods to fine-tune the hybrid parameters. This strategy has been shown to generalize well to other fine-tuning scenarios, as explored in \cite{liu2025a}. An overview of the QPA workflow is provided in Fig.~\ref{fig:qpa}.

In situations where gradient-based updates are impractical—due to noisy hardware or non-differentiable objectives—QPA remains compatible with alternative optimization strategies, such as Nelder-Mead, SPSA, or COBYLA. These non-gradient-based methods offer a robust fallback for adapting quantum-generated PEFT parameters under broader training conditions.

\subsection{QPA for Typhoon Trajectory Forecasting}


With an understanding of QPA, we can now outline our strategy for leveraging QPA to train a typhoon trajectory forecasting model using real historical typhoon trajectory data. 

\begin{figure*}[h]
\centering
\includegraphics[width=0.9\linewidth]{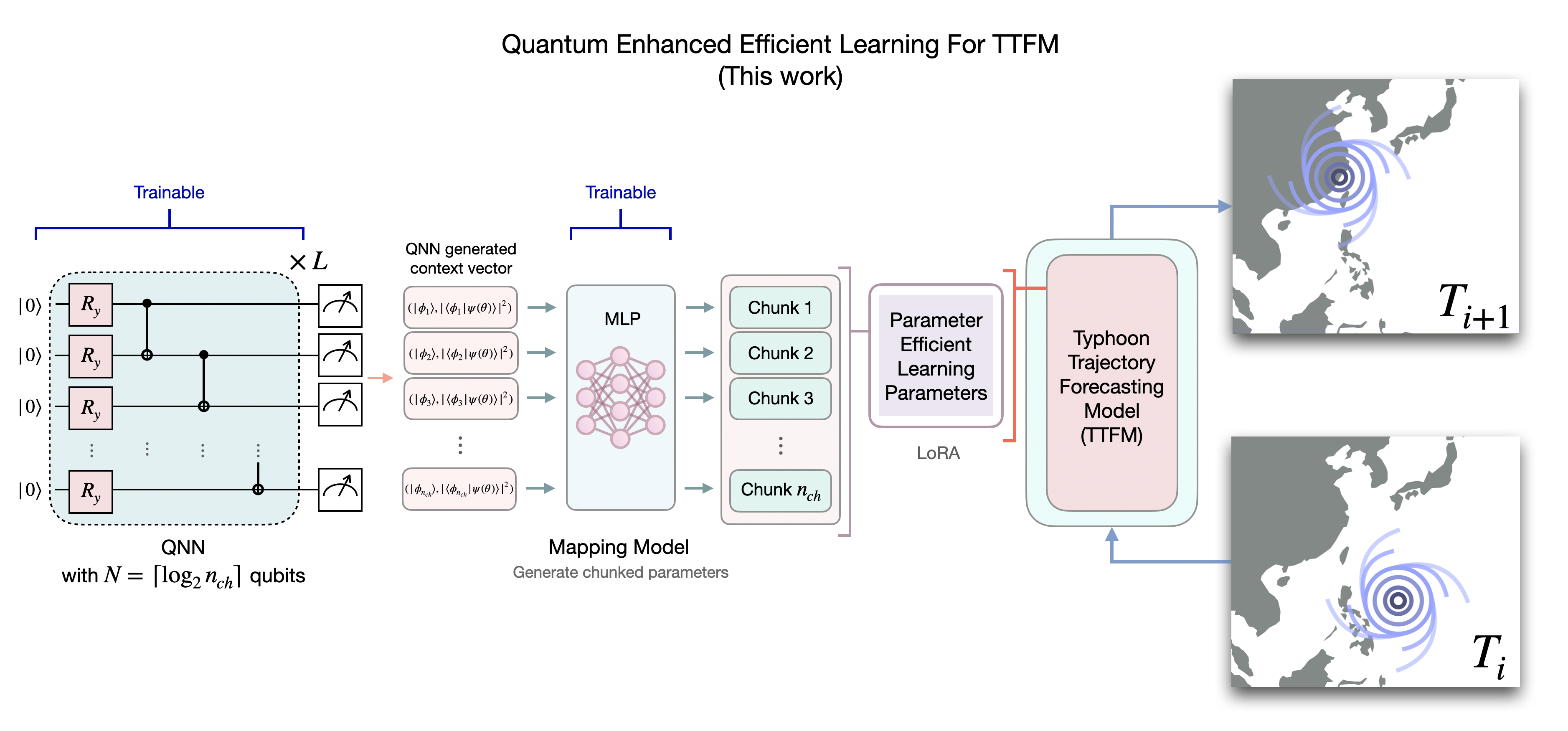}
\caption{
Overview of our solution strategy in this work.
}
 \label{fig:qeel}
\end{figure*}

To begin, we investigate classical ML approaches for typhoon trajectory forecasting and identify a prior study that achieves strong performance using an Attention-based Multi-ConvGRU model \cite{xu2022convgru}. With about 8 million parameters, this model strikes a balance between complexity and feasibility, making it an ideal candidate for enhancement through QPA within a reasonable timeframe. By applying QPA to this model, we aim to validate its effectiveness as a test case, showcasing the computational benefits and practical impact of quantum-enhanced parameter-efficient learning. Building on this foundation, we defer the details of the classical approach to the next section, while a graphical overview of our solution strategy is presented in Fig.~\ref{fig:qeel}.

\section{Data and Processing}

\begin{figure}[h]
\centering
\includegraphics[width=\linewidth]{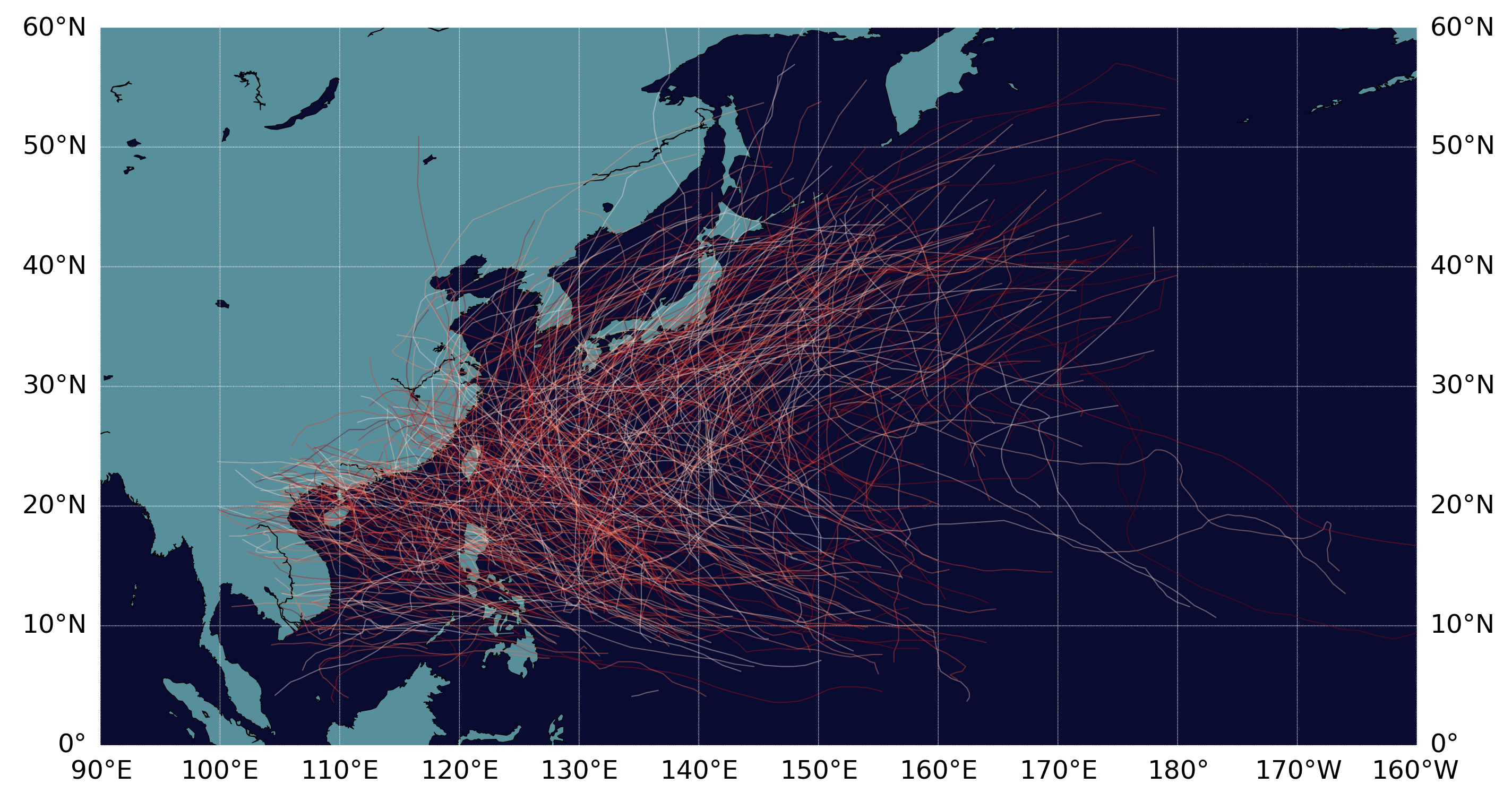}
\caption{
Visualization of typhoon trajectories from the CMA dataset.
}
 \label{fig:cma_dataset}
\end{figure}

The AM-ConvGRU model (classical target NN) is trained by QPA and evaluated using real-world typhoon datasets sourced from the China Meteorological Administration (CMA) \cite{ying2014overview} and the ERA-Interim reanalysis dataset provided by ECMWF \cite{molteni1996ecmwf}. The CMA dataset contains historical typhoon records, including 2D meteorological features such as latitude, longitude, wind speed, and central pressure. Meanwhile, the ERA-Interim dataset provides 3D atmospheric reanalysis data at multiple isobaric levels (e.g., 1000 hPa, 750 hPa, 500 hPa, 250 hPa), offering a more detailed representation of typhoon structures. The visualization of typhoon paths of the CMA dataset is shown in Fig.~\ref{fig:cma_dataset}.

To prepare the data for model training, the CMA dataset undergoes preprocessing using the CLIPER \cite{xu2022convgru} method, which extracts structured feature vectors from raw typhoon records. For the ERA-Interim dataset, the Earth’s surface is partitioned into a $1^\circ \times 1^\circ$ grid, and a $31 \times 31$ spatial window centered on the typhoon eye is selected to capture the storm’s environmental context. The RCAB module then refines the input by selecting the most relevant isobaric planes for each typhoon event. 

The model is trained using the Adam optimizer, with Mean Absolute Error (MAE) as the loss function. The Great Circle Distance metric is employed to measure the accuracy of trajectory predictions, computed as:
\begin{equation}
D = R \cos^{-1} \left[ \sin(\phi_1) \sin(\phi_2) + \cos(\phi_1) \cos(\phi_2) \cos(\lambda_2 - \lambda_1) \right],
\end{equation}
where  $\phi_1$, $\lambda_1$  and  $\phi_2$, $\lambda_2$  are the latitude and longitude coordinates of the predicted and actual typhoon locations, respectively, and  R  is the Earth’s radius.

\textbf{\textit{Training and Testing dataset.}} The AM-ConvGRU model was trained and evaluated on typhoon trajectory data spanning from 2000 to 2018, with typhoons from 2000 to 2014 used for training, and 2015 to 2018 used for testing. By leveraging both 2D meteorological features and 3D atmospheric data, AM-ConvGRU was able to reduce trajectory prediction errors in comparison to existing deep learning models like ConvLSTM. The introduction of the RCAB module proved critical in improving feature selection, particularly at different isobaric pressure levels, leading to better trajectory estimates. Moreover, due to the use of ConvGRU instead of ConvLSTM, the model achieved lower computational complexity and reduced overfitting, while maintaining high predictive performance. The Wide \& Deep framework further enhanced generalization by effectively combining statistical learning with spatio-temporal deep learning models. More classical detail on these model and method can be found in \cite{xu2022convgru}.

\section{Results}
In this section, we present the results obtained from QPA and compare them with classical model compression methods, such as pruning and weight sharing. Additionally, we analyze the performance of QPA under different hyperparameter settings, including varying chunk sizes per basis $ n_{mlp} $ and different depths of QNN layers $L$.

\subsection{Performance \& Benchmarks}
\subsubsection{Overall Benchmarking}
\begin{figure*}[h]
\centering
\includegraphics[width=0.7\linewidth]{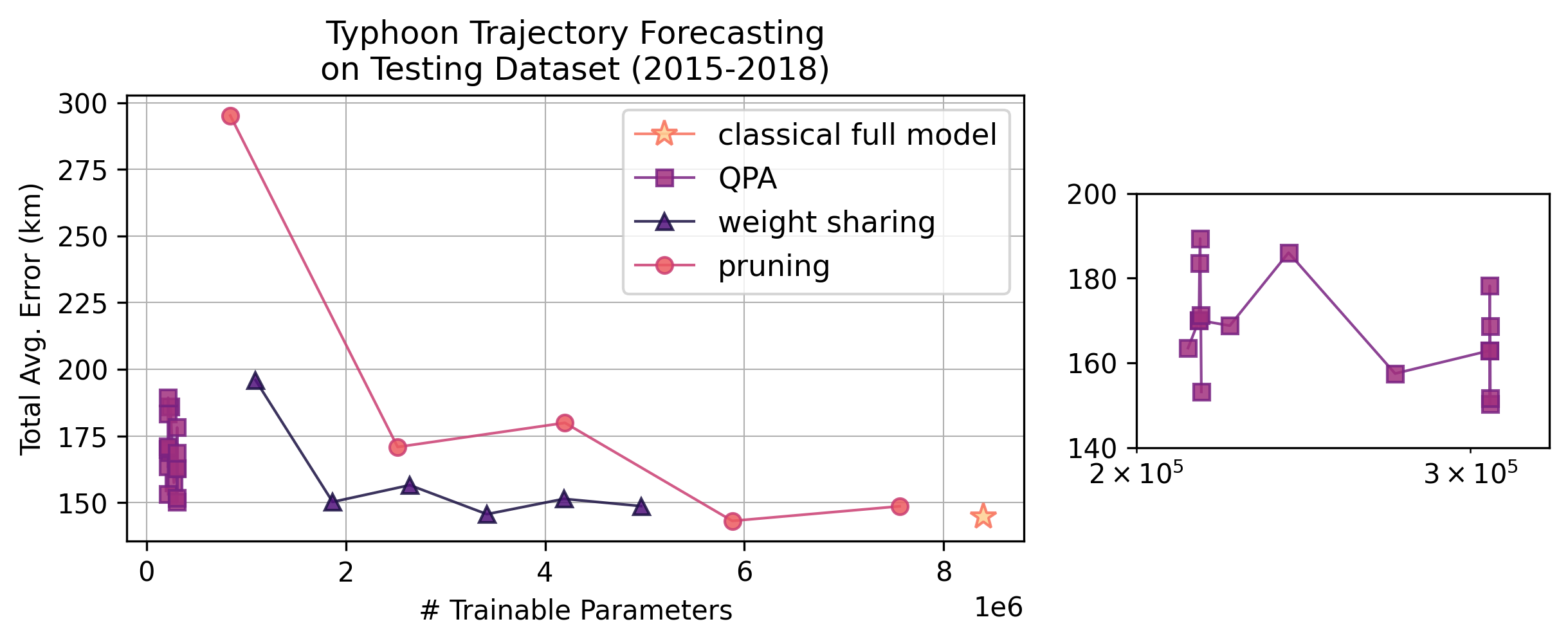}
\caption{
Typhoon trajectory forecasting on testing dataset (2015-2018).
}
 \label{fig:q_vs_cl_res}
\end{figure*}

The QPA setting in the following experiments is as follows: within the ConvGRU model for typhoon trajectory forecasting, QPA is applied to the last two linear layers, while LoRA is applied to the remaining layers. The classical full model consists of 8,399,540 (8.39M) trainable parameters. In contrast, when QPA is used with different configurations, as shown in Fig.~\ref{fig:q_vs_cl_res}, the number of trainable parameters is reduced to approximately 0.2M to 0.3M. Since QPA is fundamentally a parameter-efficient learning method that compresses the number of trainable parameters, it is crucial to compare it with other classical compression techniques, such as pruning \cite{neill2020overviewneuralnetworkcompression, blalock2020state} and weight sharing \cite{nowlan2018simplifying}.

Fig.~\ref{fig:q_vs_cl_res} illustrates the performance comparison of different model compression methods applied to typhoon trajectory forecasting on a testing dataset spanning 2015 to 2018. The x-axis represents the number of trainable parameters, while the y-axis denotes the total average error (km) for the forecasting task, where lower values indicate higher forecasting accuracy. The objective of this study is to assess the effectiveness of various compression techniques, including QPA, weight sharing, and pruning, in reducing model size while maintaining predictive accuracy.

The classical full model, as proposed in the original paper \cite{xu2022convgru}, is represented by an orange star and achieves the lowest forecasting error but requires significantly more trainable parameters, exceeding 8.39M. While larger models tend to achieve higher accuracy, their computational and energy costs can be prohibitive. Pruning, represented by red circles, demonstrates a steep decline in error as the number of trainable parameters decreases. However, after reaching approximately $10^6$ parameters, the reduction in error plateaus, suggesting diminishing returns from further pruning.

The weight sharing method, depicted by dark blue triangles, offers a more balanced trade-off between parameter reduction and forecasting accuracy. Compared to pruning, weight sharing maintains relatively stable accuracy with significantly fewer trainable parameters, indicating that it is an effective strategy for improving model efficiency. Meanwhile, QPA, shown as purple squares, consistently achieves competitive performance with a significantly lower parameter count. The clustering of QPA results on the left side of the plot within the 0.2M to 0.3M parameter range highlights its potential to maintain accuracy while substantially reducing computational requirements.
The inset plot provides a magnified view of QPA’s performance, specifically within the 0.2M to 0.3M parameter range. While minor fluctuations in total average error are observed, the results remain within a narrow band, reinforcing QPA’s robustness even under significant parameter compression. This suggests that QPA provides a viable alternative to traditional compression methods by leveraging quantum techniques to optimize model efficiency. The content of this inset plot is expanded in Fig.~\ref{fig:qpa_diff_prop}.
The results suggest that QPA, a quantum method leveraging quantum principles, offers a promising alternative to classical full models by representing high-dimensional complex data with fewer trainable parameters. By utilizing quantum state representations mappings, QPA efficiently captures intricate data structures, enabling substantial reductions in model size while preserving predictive accuracy. While pruning proves effective at higher parameter levels, it reaches a performance plateau beyond a certain threshold, limiting its scalability. In contrast, QPA’s ability to compress information using quantum techniques allows it to maintain competitive forecasting accuracy while significantly lowering computational complexity. These findings underscore the potential of QPA as an energy-efficient quantum-enhanced approach for machine learning applications, offering a scalable and resource-efficient solution for real-world forecasting tasks.

\subsubsection{On Hyperparameters of QPA}
\begin{figure*}[h]
\centering
\includegraphics[width=0.7\linewidth]{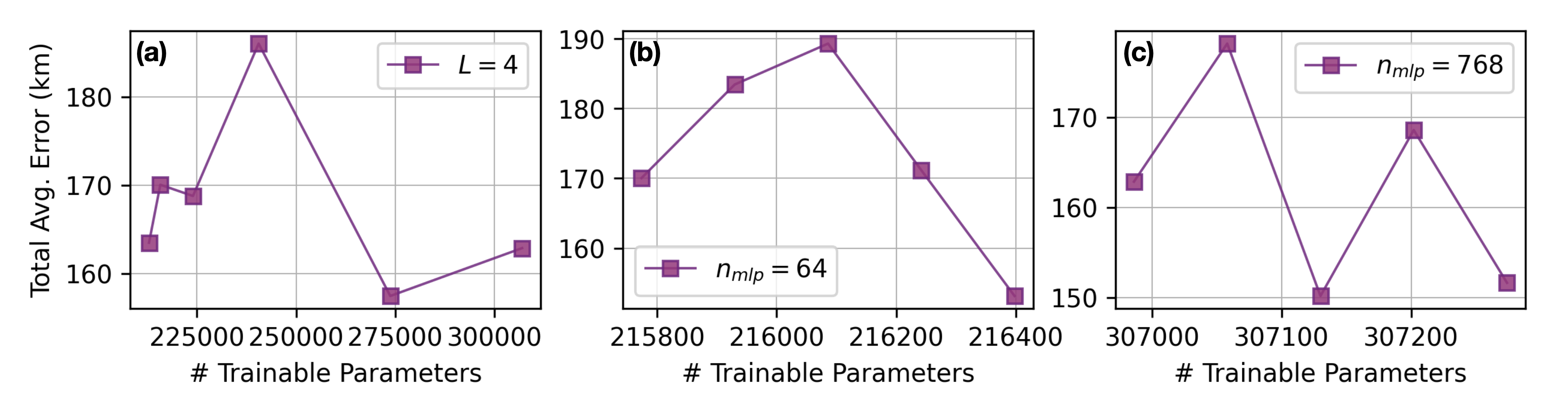}
\caption{
Typhoon trajectory forecasting on testing dataset (2015-2018). (a) Fixing \# of QNN layers $L = 4$, the corresponding required qubit size from left to right is 9, 8, 7, 6, 5, 4.  (b) and (c): fixing $n_{mlp}$ and varying the QNN depth $L = 4, 8, 12, 16, 20$. 
}
 \label{fig:qpa_diff_prop}
\end{figure*}

Fig.~\ref{fig:qpa_diff_prop} consists of three subplots, each examining the impact of different QPA hyperparameters on forecasting accuracy. The general trend observed across all subplots suggests that increasing the number of trainable parameters tends to improve forecasting performance, resulting in lower error values. Fig.~\ref{fig:qpa_diff_prop}(a), the QNN layer depth is fixed at  $L = 4 $, while the chunk size $ n_{mlp}$  is varied across the values 32, 64, 128, 256, 512, and 768. The results indicate that the total average error fluctuates across different chunk sizes, but there is no strictly monotonic trend. However, configurations with higher trainable parameter counts generally achieve lower errors, reinforcing the parameter-performance trade-off in quantum models.

\begin{figure}[h]
\centering
\includegraphics[width=\linewidth]{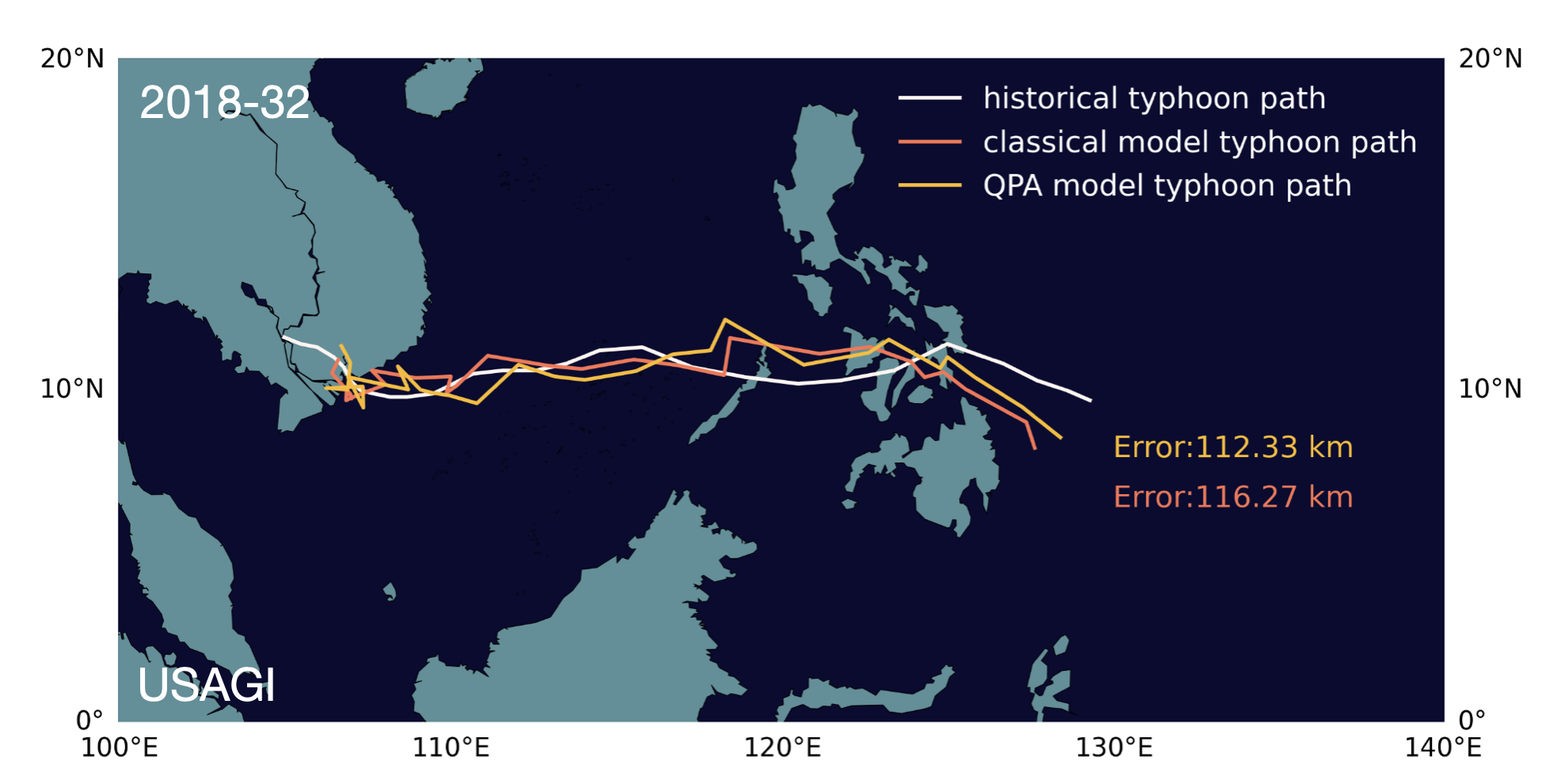}
\caption{
Typhoon trajectory forecasting results for typhoon USAGI, \# 2018-32. 
}
 \label{fig:usagi_traj}
\end{figure}

In Fig.~\ref{fig:qpa_diff_prop}(b), the chunk size is fixed at $ n_{mlp} = 64 $, while the number of QNN layers is varied over $ L = 4, 8, 12, 16, 20 $. The results exhibit a similar trend, where increasing QNN depth initially leads to higher error values, possibly due to optimization difficulties in shallow models. However, as the number of trainable parameters increases, the total average error decreases, suggesting that a deeper QNN architecture improves prediction accuracy. Fig.~\ref{fig:qpa_diff_prop} (c), the experiment is identical to (b) but with a larger chunk size fixed at  $n_{mlp} = 768 $. A similar pattern emerges, where models with more trainable parameters achieve lower errors, reinforcing the idea that larger models can capture more complex features and improve generalization.

\subsubsection{Resulting Trajectories Visualization}
To better understand the performance of QPA, it is beneficial to visualize the predicted typhoon trajectories. The ability to accurately forecast typhoon paths is crucial for early warning systems, as it enables timely evacuations, reduces financial losses, and, most importantly, \textbf{saves lives}. The effectiveness of a forecasting model is not only measured by numerical accuracy but also by how well it captures the essential patterns and deviations in real-world trajectories.

\begin{figure}[h]
\centering
\includegraphics[width=\linewidth]{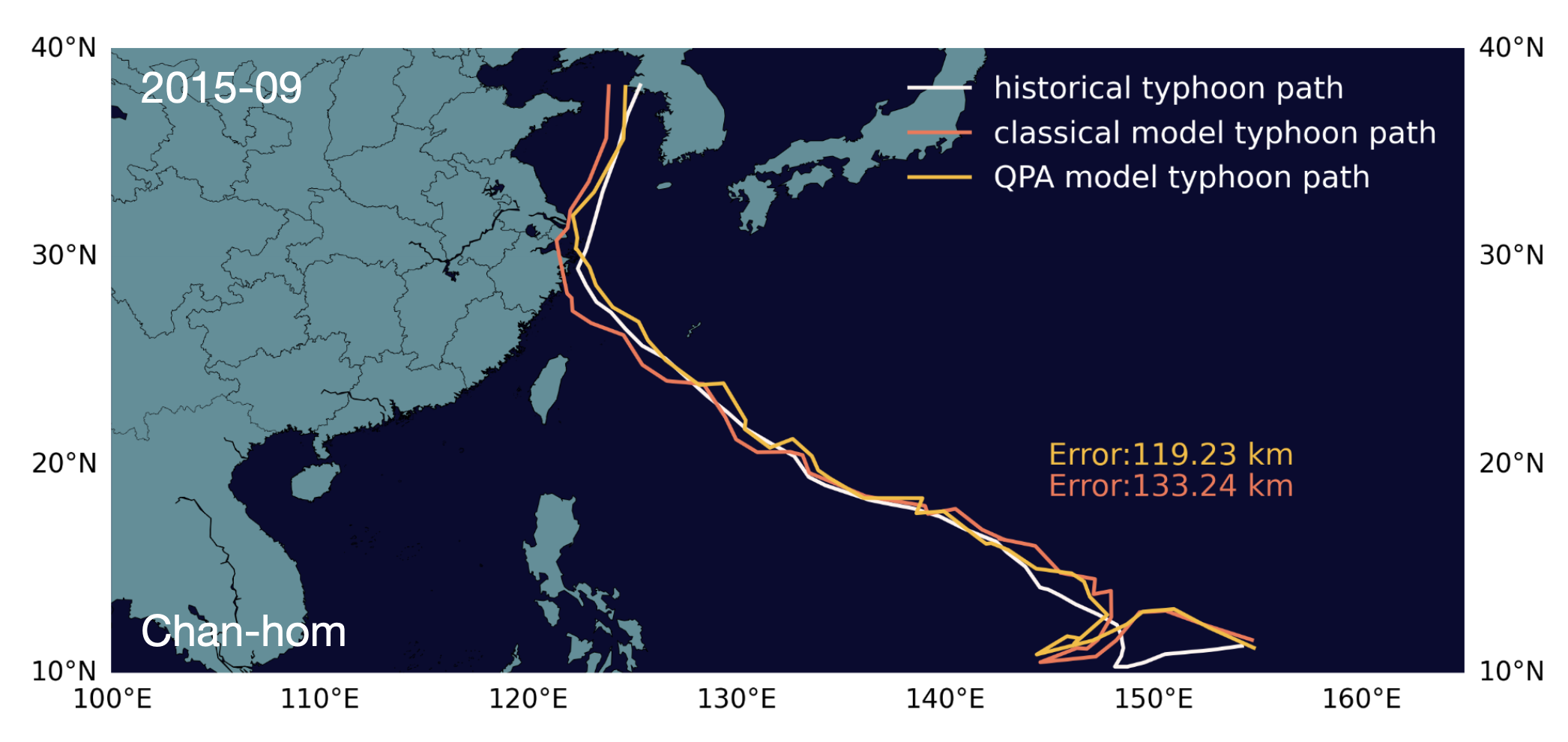}
\caption{
Typhoon trajectory forecasting results for typhoon Chan-hom, \# 2015-09. 
}
 \label{fig:chanhom_traj}
\end{figure}

\begin{figure}[h]
\centering
\includegraphics[width=\linewidth]{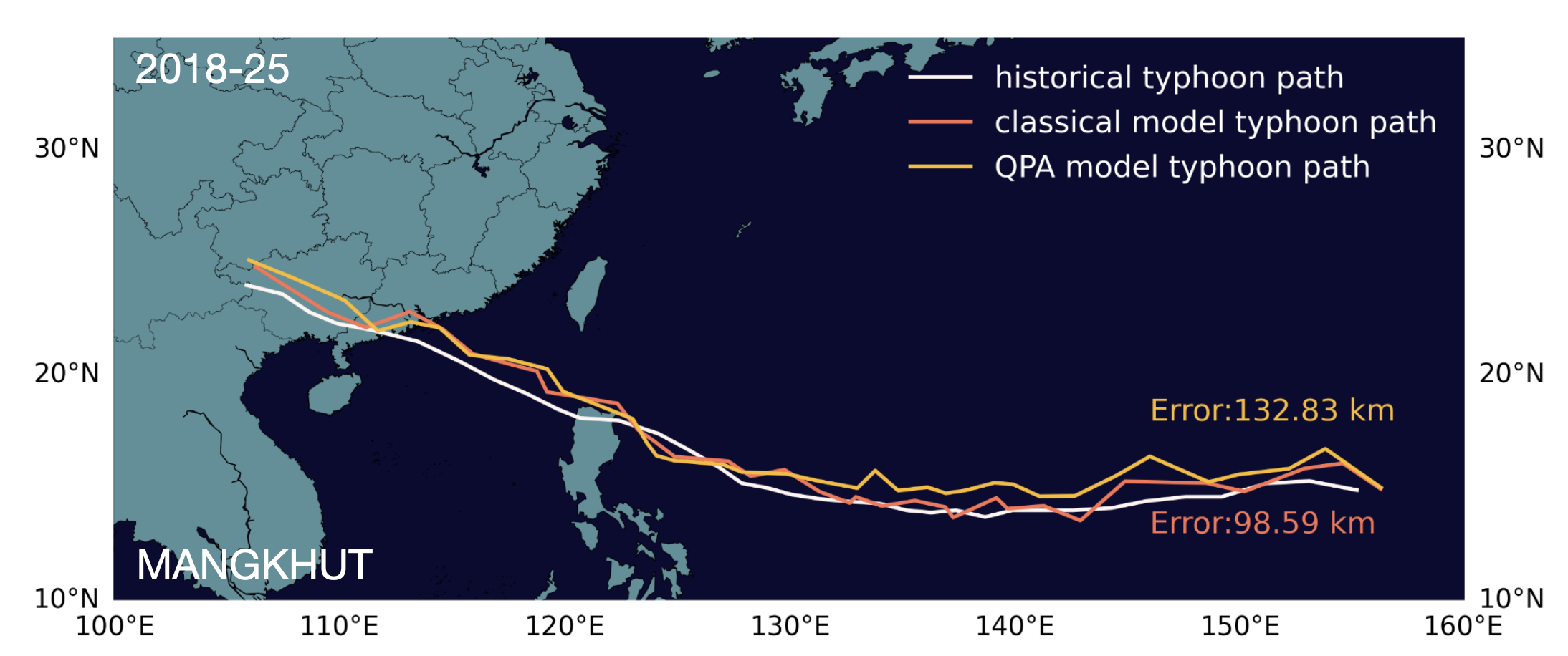}
\caption{
Typhoon trajectory forecasting results for typhoon MANGKHUT, \# 2018-25. 
}
 \label{fig:mangkhut_traj}
\end{figure}

\begin{figure}[h]
\centering
\includegraphics[width=\linewidth]{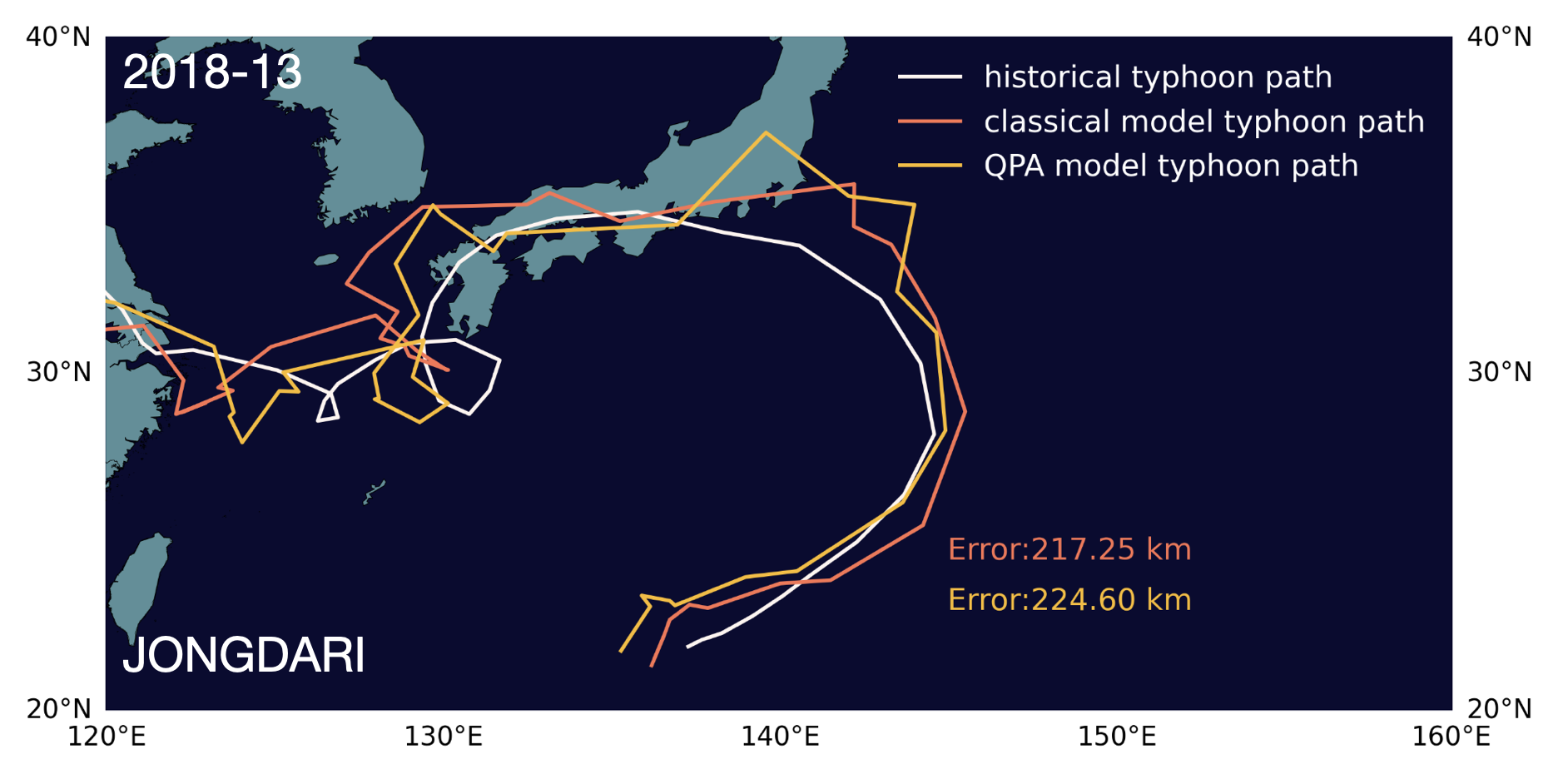}
\caption{
Typhoon trajectory forecasting results for typhoon JONGDARI, \# 2018-13. 
}
 \label{fig:jongdari_traj}
\end{figure}

In Fig.~\ref{fig:usagi_traj} and Fig.~\ref{fig:chanhom_traj}, we present two cases where QPA, with approximately 0.3M trainable parameters, outperforms the original model with 8.3M trainable parameters. These results demonstrate that QPA, despite using only a fraction of the trainable parameters, effectively captures complex atmospheric dynamics and provides trajectory forecasts that closely align with the actual typhoon paths. It is particularly impressive that even with a significantly reduced parameter count—nearly dozens of times smaller than the full model—the deviation in error remains minimal. This suggests that QPA’s quantum-inspired parameterization efficiently encodes critical weather patterns, making it a promising approach for resource-efficient and scalable typhoon prediction. Conversely, Fig.~\ref{fig:mangkhut_traj} and Fig.~\ref{fig:jongdari_traj} illustrate cases where QPA does not surpass the original full model. While QPA still provides reasonably accurate predictions, these results highlight the limitations of parameter-efficient models in capturing extreme meteorological variations, which may require larger network capacity to fully resolve. Nevertheless, the fact that QPA maintains competitive performance with significantly fewer parameters suggests that it is a viable approach for improving forecasting efficiency. These instances indicate that while QPA is an effective method for reducing computational costs, further refinements—such as hybrid quantum-classical architectures or adaptive parameterization techniques—may enhance its performance in more complex forecasting scenarios.

\subsubsection{Compatibility}
A key aspect of deployment is ensuring compatibility between QPA and quantum processors, there is a clear pathway for integrating QPA into near-term quantum systems. Additionally, ongoing research in error mitigation strategies and hardware-aware optimizations will further refine QPA-based learning, ensuring it remains robust against quantum noise and decoherence effects.
Integrating QPA into existing high-performance computing (HPC) environments represents another exciting frontier. Most large-scale weather forecasting models, including typhoon trajectory prediction systems, are designed for classical supercomputing infrastructure. The transition to a hybrid quantum-classical framework will unlock new opportunities for faster, more energy-efficient computations, provided that efficient interfacing techniques are developed. By optimizing data flow and parallel execution between quantum and classical processors, large-scale deployments can seamlessly leverage quantum advantages while maintaining compatibility with existing computational pipelines.

\section{Conclusion}

In this study, we introduced the application of \textit{Quantum Parameter Adaptation} (QPA) to the task of typhoon trajectory forecasting, demonstrating its practical effectiveness in a real-world climate modeling scenario. By applying the QPA framework with a LoRA-style PEFT scheme, we successfully reduced the number of trainable parameters from 8.39 million in the original full model to just 0.3 million—a reduction of over 96\%. Remarkably, this highly compressed model achieved forecasting performance comparable to the full baseline model.

Moreover, QPA was shown to outperform classical model compression techniques such as pruning and weight sharing, both in terms of prediction accuracy and generalization, while maintaining a significantly smaller parameter footprint. These findings highlight QPA’s advantage as a hybrid quantum-classical method that not only preserves but enhances performance under extreme compression.

This study underscores the potential of QPA as a scalable, efficient, and sustainable solution for large-scale geoscientific applications. As quantum hardware continues to mature, the use of QPA in climate forecasting and other data-intensive domains may unlock new frontiers in energy-efficient machine learning.  While the findings in this report demonstrate the feasibility of QPA using simulated quantum systems, further research and development will be necessary to unlock its full potential. Future efforts should focus on refining hybrid architectures, improving measurement efficiency, optimizing quantum learning models, and contributing to the sustainability of AI-driven climate forecasting.

\subsection{Advancing Hybrid Quantum-Classical Integration}
A key avenue for future exploration lies in the continued development of hybrid quantum-classical learning strategies. Optimizing QPA through techniques such as continuous-time Hamiltonian evolution, variational quantum algorithms, and quantum-native optimization could reduce reliance on purely gate-based approaches, improving training efficiency. Additionally, the incorporation of hardware-aware error mitigation strategies will be essential for stable and scalable learning on real quantum devices. Future research should prioritize the development of seamless hybrid training pipelines, where quantum-generated parameter updates interact efficiently with classical optimization processes to enhance both model accuracy and resource efficiency.

\subsection{Beyond Climate Forecasting}
Beyond technical improvements, QPA’s potential role in promoting sustainable AI practices is another promising direction. As energy efficiency becomes a growing concern in the training of large-scale models, QPA-based learning offers the possibility of significantly reducing the carbon footprint associated with full-model classical training. This efficiency gain can be extended beyond climate forecasting, with potential impact on resource-intensive AI applications such as natural language processing and financial modeling. Collaborative efforts between quantum computing researchers, AI developers, and sustainability experts will be crucial for building quantum-assisted learning systems aligned with environmental objectives.

In the long term, integrating QPA into next-generation quantum computing platforms will be critical in shaping the future of hybrid AI systems. As quantum processors scale in size and reliability, QPA's adaptability and model compression capabilities will play a key role in advancing parameter-efficient learning. Continued research on algorithmic optimization and hardware-aware model design will ensure that QPA remains a leading approach for building practical, energy-conscious AI systems capable of addressing real-world challenges.

\section*{Acknowledgment}
Wei-Hao Huang (Jij inc) acknowledges this work was performed for the Council for Science, Technology and
Innovation (CSTI), Cross-ministerial Strategic Innovation Promotion Program (SIP), ``Promoting the application of advanced quantum technology platforms to social issues'' (Funding agency: QST).
This work was supported by the Engineering and Physical Sciences Research Council (EPSRC) under grant number EP/W032643/1.

\clearpage
\bibliographystyle{IEEEtran}
\bibliography{bib/tools,bib/vqc,bib/qml_examples,bib/quantum_fl, bib/ml_examples, bib/hybrid_co_examples,bib/classical_fl,references,bib/fwp,bib/qt}

\end{document}